\documentclass{egpubl}
\usepackage{egsgp2025}

\SpecialIssuePaper

\usepackage[T1]{fontenc}
\usepackage{dfadobe}  

\usepackage{cite}  % comment out for biblatex with backend=biber
\BibtexOrBiblatex
\electronicVersion
\PrintedOrElectronic
\ifpdf \usepackage[pdftex]{graphicx} \pdfcompresslevel=9
\else \usepackage[dvips]{graphicx} \fi

\newcommand{\method}{~Im2SurfTex} %< I changed something and 
\renewcommand{\paragraph}[1]{\vspace{1em}\noindent\textbf{#1}.}

\usepackage{amsmath}
\usepackage{amssymb}
\usepackage{booktabs}
\usepackage{xcolor}
\usepackage{overpic}
\usepackage{enumitem} %< control spacing in itemize/enumerate/...
\usepackage{overpic} %< add raw math symbols to figures
\usepackage{color}
\usepackage{colortbl}
\usepackage{multirow}
\usepackage[ruled]{algorithm2e}
\RestyleAlgo{ruled}
\usepackage{comment}
\usepackage{wrapfig}
\usepackage[export]{adjustbox}
\usepackage{manyfoot}
\SetFootnoteHook{\hspace*{-0.5em}}
\DeclareNewFootnote{bl}[gobble]
\usepackage{relsize}

\definecolor{turquoise}{cmyk}{0.65,0,0.1,0.3}
\definecolor{purple}{rgb}{0.65,0,0.65}
\definecolor{dark_green}{rgb}{0, 0.5, 0}
\definecolor{orange}{rgb}{0.8, 0.6, 0.2}
\definecolor{red}{rgb}{0.8, 0.2, 0.2}
\definecolor{darkred}{rgb}{0.6, 0.1, 0.05}
\definecolor{blueish}{rgb}{0.0, 0.3, .6}
\definecolor{light_gray}{rgb}{0.7, 0.7, .7}
\definecolor{pink}{rgb}{1, 0, 1}
\definecolor{greyblue}{rgb}{0.25, 0.25, 1}

\DeclareMathOperator*{\avg}{avg}

\newcommand{\real}{\mathbb{R}}

\newcommand{\bc}{\mathbf{c}}

\newcommand{\bff}{\mathbf{f}}

\newcommand{\bh}{\mathbf{h}}

\newcommand{\bk}{\mathbf{k}}

\newcommand{\bn}{\mathbf{n}}

\newcommand{\bp}{\mathbf{p}}
\newcommand{\bq}{\mathbf{q}}

\newcommand{\bs}{\mathbf{s}}
\newcommand{\bt}{\mathbf{t}}
\newcommand{\bu}{\mathbf{u}}
\newcommand{\bv}{\mathbf{v}}

\newcommand{\bz}{\mathbf{z}}

\newcommand{\bD}{\mathbf{D}}

\newcommand{\bG}{\mathbf{G}}

\newcommand{\bI}{\mathbf{I}}

\newcommand{\bK}{\mathbf{K}}

\newcommand{\bQ}{\mathbf{Q}}

\newcommand{\bS}{\mathbf{S}}
\newcommand{\bT}{\mathbf{T}}

\newcommand{\bV}{\mathbf{V}}

\newcommand{\mR}{\mathcal{R}}
\newcommand{\mD}{\mathcal{D}}

\newcommand{\mN}{\mathcal{N}}

\usepackage{egweblnk} 
\title[\method]%
      {\method: Surface Texture Generation via Neural Backprojection of Multi-View Images}
      
\author[Y.\ Georgiou et al.]
{\parbox{\textwidth}{\centering\vspace*{-7mm}
    Yiangos Georgiou$^{1,2}$\orcid{0000-0001-9604-3301}, 
    Marios Loizou$^{1,2,3}$\orcid{0000-0002-2920-0087}, 
    Melinos Averkiou$^{1,2}$\orcid{0000-0003-1814-7134} and 
    Evangelos Kalogerakis$^{2,3}$\orcid{0000-0002-5867-5735}}
        \\
{\parbox{\textwidth}{\centering\vspace*{-7mm}
    $^1$University of Cyprus ~~~~ 
    $^2$CYENS CoE ~~~~ 
    $^3$Technical University of Crete}
}
}
\begin{document}

% uncomment for using teaser
\teaser{
   \vspace*{-12mm}
   \includegraphics[width=\textwidth]{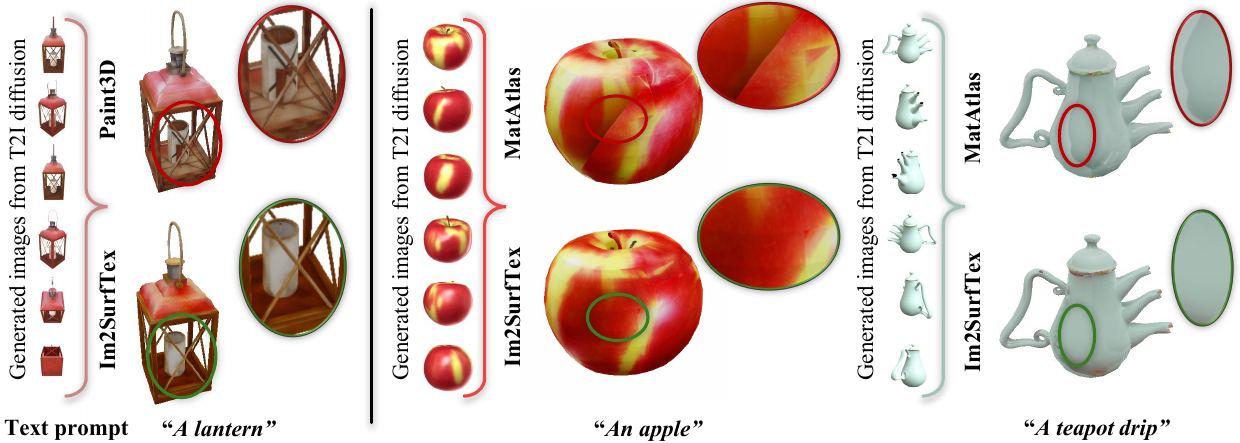}
    \centering
    \vspace*{-6mm}
    \caption{Given a text prompt and an untextured 3D shape,\method \ generates a texture 
    for it by learning to backproject images produced by text-to-image (T2I) diffusion 
    models to the shape's texture space. \emph{Left:} \method \ diminishes artifacts on 
    surfaces with self-occlusions and complex geometry, preserving finer details where 
    alternatives like Paint3D~\cite{Zeng:2024:Paint3D} struggle, resulting in 
    backprojection issues, such as the guard grill's texture appearing on the candle inside 
    the lantern. \emph{Right:} \method \ prevents seam formation on high-curvature surfaces 
    and seamlessly blends multiple views. In contrast, other approaches, such as 
    MatAtlas~\cite{Ceylan:2024:Matatlas}, often introduce texture discontinuities, as seen 
    on the apple, or fail to resolve multi-view inconsistencies, leading to visible 
    artifacts, as in the teapot.}
    \label{fig:teaser}
}

\maketitle

%%%% Abstract
\vspace*{-7mm}
\begin{abstract}
We present\method, a method that generates textures for input 3D shapes by learning to aggregate multi-view image outputs produced by 2D image diffusion models onto the shapes' texture space.  Unlike existing texture generation techniques that use ad hoc backprojection and averaging schemes to blend multiview images into textures, often resulting in texture seams and artifacts, our approach employs a trained neural module to boost texture coherency.
 The key ingredient of our module is to leverage neural attention and appropriate positional encodings of image pixels based on their corresponding 3D point positions, normals, and surface-aware coordinates as encoded in geodesic distances within surface patches.  These encodings capture texture correlations between neighboring surface points, ensuring better texture continuity. 
 Experimental results show that our module improves texture quality, achieving superior performance in high-resolution texture generation.

%-------------------------------------------------------------------------
%  ACM CCS 1998
%  (see http://www.acm.org/about/class/1998)
% \begin{classification} % according to http:http://www.acm.org/about/class/1998
% \CCScat{Computer Graphics}{I.3.3}{Picture/Image Generation}{Line and curve generation}
% \end{classification}
%-------------------------------------------------------------------------
%  ACM CCS 2012
% (see http://www.acm.org/about/class/class/2012)
%The tool at \url{http://dl.acm.org/ccs.cfm} can be used to generate
% CCS codes.


\end{abstract}

\vspace*{-2mm}
%%%% Introduction
\section{Introduction}
\label{sec:intro} 

\begin{figure*}[!t]
    \includegraphics[width=\textwidth]{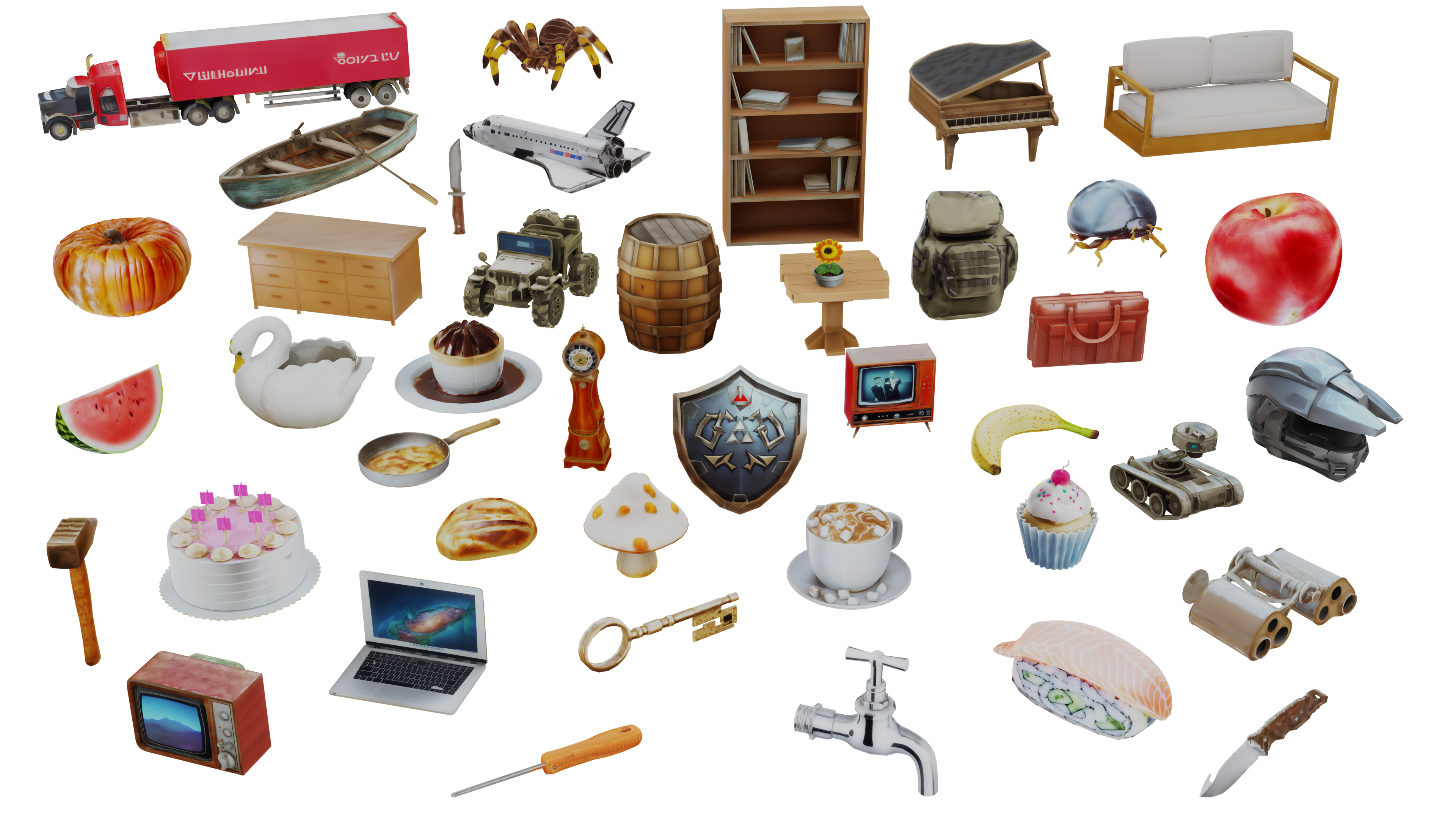}
    \vspace*{-7mm}
    \caption{A gallery of 3D shapes from various categories, textured by \method.}
    \label{fig:gallery}
    \vspace*{-5mm}
\end{figure*}

\footnotebl{$^{\mathlarger{\dagger}} \ \ $In accordance with the ERC Open Access mandate, the authors have made the Author Accepted Manuscript (AAM) publicly available under the Creative Commons Attribution (CC-BY 4.0) license.}

\vspace*{-4mm}
Producing compelling 3D assets has become an increasingly active area of research in the field of generative AI. 
Despite the significant progress in training large-scale generative models of 3D geometry~\cite{Lin:2023:Magic3D, 
Liu:2023:Zero, Wang:2024:Prolificdreamer, Gao:2022:Get3d, Jun:2023:Shape, Νichol:2022:Pointe, Vahdat:2022:Lion, 
Liu:2024:One, Liu:2023:One2345plus, Shi:2023:Zero123plus, Zhang:2024:CLAY}, synthesizing compelling, seamless, and 
high-quality textures for 3D shapes and scenes still remains challenging. One major obstacle is the limited 
availability of training 3D asset datasets with high-quality textures. Several recent 
approaches~\cite{Chen:2023:Text2tex, Richardson:2023:Texture, 
Cao:2023:Texfusion, Ceylan:2024:Matatlas, Zeng:2024:Paint3D, Liu:2024:SyncMVD, Cheng:2024:MVPaint}  have resorted 
to leveraging powerful generative 2D image models, such as denoising diffusion models pre-trained on massive 2D 
datasets, to guide surface texture generation. These approaches generate images across different views conditioned 
on text prompts and depth maps rendered from the given 3D shapes or scenes, then attempt to combine the generated 
images onto the surface.

Unfortunately, these methods suffer from a number of limitations. First, merely projecting the generated images 
back to the surface tends to generate texture distortions in areas of rapid depth changes or high curvature
surface regions (Figure \ref{fig:teaser}, left). Second, the images generated from different views are often combined 
with ad-hoc criteria to create a single surface texture. For example, many methods~\cite{Chen:2023:Text2tex, Richardson:2023:Texture}
simply transfer the colors from the most front-facing view to each texel (i.e., pixel in the texture map), causing 
visible seams in the texture maps especially in areas where colors of neighboring texels are copied from different 
views (Figure \ref{fig:teaser}, middle \& right). Other methods rely on simple averaging schemes of RGB 
colors~\cite{Ceylan:2024:Matatlas} or image latents, causing significant color bleeding. 
Others resort to global texture optimization techniques
\cite{Cao:2023:Texfusion, Liu:2024:SyncMVD}, which are slow and can still fail to generate 
coherent textures since their initialization is still based on ad-hoc thresholds for view 
selection and color averaging. 

Our approach, named \method, addresses the above limitations with the introduction of a novel, optimization-free module 
that can be easily integrated to existing texture generation approaches. The module is trained to combine color 
information from multiple  viewpoints to textures through a cross-attention mechanism, where for each surface 
point, several candidate image neighborhoods across different views are examined and combined back in texture 
space depending on local surface geometry, 
as encoded in 3D positions, normals and geodesic distances within these patches. 
In this manner, the attention mechanism captures local context based 
on surface (geodesic) proximity rather than relying solely on 3D Euclidean proximity that might correlate surface region textures far from each other 
in geodesic sense. Our module yields coherent textures efficiently, without requiring any slow optimization procedures. Our experiments indicate significant improvements in generated texture quality, measured by different scores, including 
FID~\cite{Heusel:2017:GANs}, KID~\cite{Bińkowski:2018d:Demystifying}, CLIP~\cite{Schuhmann:2022:Laion5b}
metrics, when compared to alternatives. 

In summary, our method introduces the following contributions:
\begin{itemize}
    \item a cross-attention mechanism that learns how to wrap generated images from different views onto a single, coherent surface texture map.
    \item we integrate this modules with multiple alternative backbones based on 
	texture map diffusion showing consistent improvements in synthesized texture quality.
\end{itemize}

%%%% Related work
\section{Related Work}

\paragraph{Early Works on Texture Synthesis}
Classic texture generation methods mostly focused on example-based approaches~\cite{Wei:2009:STARtexture}, 
including region-growing techniques~\cite{Efros:1999:texture, Wei:2000:Fast} and strategies leveraging local
coherence~\cite{Ashikhmin:2001:Synthesizing, Tong:2002:BTF} to maintain consistency across synthesized 
textures. Patch-based methods~\cite{Efros:2001:Quilting, Kwatra:2003:Graphcuttex} 
synthesized textures using patch patterns from reference images, while other
techniques~\cite{Kwatra:2005:TextureOptim, Han:2006:FastOptim} progressively refined 
synthesized texture based on optimization procedures. Another significant research direction involved directly generating textures on 3D 
surfaces~\cite{Turk:2001:Surfaces, Zhang:2006:vector, Fisher:2007:Design} by exploiting vector 
fields defined over the surface to seamlessly map textures onto complex geometries. All these example-based approaches were unable to capture texture variability, generate diverse textures, or handle diverse shapes.

\paragraph{Text-Guided Diffusion Models for Image Synthesis}
Our method builds upon diffusion models~\cite{Sohl:2015:Deep, Ho:2020:Denoising, Zhang:2023:T2I}, which have demonstrated superior performance compared to GANs~\cite{Goodfellow:2014:GAN, Zhu:2017:unpaired} in image generation tasks~\cite{Rombach:2022:High, Nichol:2021:glide, Saharia:2022:imagen, Ramesh:2022:dalle, 
Zhang:2023:ControlNet}. 
Closely related to our approach are text-guided generation models for 3D object synthesis, 
where text-to-image diffusion models are used for distilling 3D objects as neural radiance 
fields~\cite{Mildenhall:2020:nerf, Kerbl:2023:3Dgaussians} via Score Distillation 
Sampling~\cite{Poole:2022:Dreamfusion, Wang:2023:score}. Following DreamFusion~\cite{Poole:2022:Dreamfusion}, several approaches have been proposed~\cite{Lin:2023:Magic3D, 
Metzer:2023:latentnerf, Chen:2024:text2gs, Wang:2024:Prolificdreamer, Tsalicoglou:2024:textmesh, 
Shi:2024:mvdream}. However, these methods do not specifically target the task of texture generation.

\paragraph{Texture Generation via T2I Diffusion Models}
Initial efforts in texture generation via text-to-image diffusion models, such as 
Text2Tex~\cite{Chen:2023:Text2tex} and TEXTure~\cite{Richardson:2023:Texture}, employed depth-conditioned 
diffusion models~\cite{Rombach:2022:High, Zhang:2023:ControlNet} to iteratively inpaint and refine the 
textures of 3D objects. Both methods start with a preset viewpoint, generating texture updates for 
corresponding regions of the 3D object by back-projecting depth-guided views. In Text2Tex, a coarse texture 
is progressively created by iterating over multiple viewpoints and refining the texture map based on high-
surface-coverage viewpoints. This refinement applies a denoising diffusion process of moderate strength to 
preserve the texture’s original appearance while enhancing details. Similarly, TEXTure divides the texture 
map into distinct regions labeled as \textit{keep}, \textit{refine}, or \textit{generate}, enabling 
selective refinement or generation of textures. Despite these efforts to achieve global consistency, these methods employ ad hoc thresholds to define the different shape regions and hand-engineered strategies for backprojection, often leading to seams between texture regions synthesized from different viewpoints.

To address these issues, other methods such as TexFusion~\cite{Cao:2023:Texfusion} leverage 
\textit{latent} diffusion to interlace diffusion and back-projection steps, producing 3D-aware 
latent images that are subsequently decoded and merged into a texture map. Similarly, 
SyncMVD~\cite{Liu:2024:SyncMVD} employs a latent texture map where all views are encoded at each 
denoising step, further enhancing consistency in geometry and appearance. TexGen~\cite{Huo:2024:TexGen} introduced a multi-view sampling and resampling framework 
that updates a UV texture map iteratively during denoising, aiming to reduce view discrepancies.
Still, these methods rely on ad hoc blending masks or heuristics for aggregating texture information from different views, such as mere averaging or using the most front-facing view information for each texel.
 In contrast, our approach learns to aggregate color information from multiple views based on both geometry and texture information, promoting the generation of more coherent and seamless surface texture maps.

A more recent approach, Paint3D~\cite{Zeng:2024:Paint3D}, achieves impressive texture generation results, by adopting a two-stage texture generation strategy. 
In the first stage, a coarse texture is created by backprojecting views to texture space via the heuristic of
using the most front-facing view information for each texel, as in previous methods. The second stage involves a 
refinement and inpainting process that utilizes a diffusion model in UV texture space, conditioned on a UV position map 
encoding the 3D adjacency information of texels. While this method directly encodes 3D geometric 
information into the texture map, it can still result in misaligned textures due to its employed heuristic during its coarse stage. 
The subsequent texture refinement steps often fail to fix the artifacts of the coarse stage, as demonstrated in our experiments.
Another related method, MatAtlas~\cite{Ceylan:2024:Matatlas}, incorporates a three-step denoising process 
with sequential operations and line conditions to preserve geometry and style consistency. However, 
MatAtlas employs an averaging heuristic for blending the final texture from the generated views, leading to inconsistencies or overly smooth surfaces, as also demonstrated in our experiments. TEXGen~\cite{Yu:2024:TEXGen} takes a different approach by directly training a large-scale diffusion model in the UV texture 
space, and integrating convolution operations 
in UV space with 3D-aware attention layers in their denoising network to achieve high-resolution texture synthesis. However, their method still faces challenges in maintaining cross-view consistency since the 
generated textures are conditioned on single-view images which are merely backprojected to the UV space to derive the initial partial texture maps used in their diffusion.

In a concurrent work, MVPaint~\cite{Cheng:2024:MVPaint} introduces a multi-stage texture generation framework. In the 
initial stage, a latent texture map is employed during multi-view projection to create a synchronized texture 
across multiple views, similar to SyncMVD~\cite{Liu:2024:SyncMVD}. This is followed by an inpainting stage, where 
uncovered texture regions are filled using a dense colored point cloud extracted from the generated texture map. 
Colors are propagated to empty texels in a spatially aware manner using inverse distance weighting and normal 
similarity between neighboring points. Finally, a refinement stage upscales the texture map and smooths out seams 
through weighted color averaging among k-nearest neighbors in 3D space. Still, MVPaint relies on an averaging scheme to
aggregate view information into texture space. In contract, our approach learns this aggregation by encoding both 
geometric and appearance information from multiple views to produce textures with greater  consistency.

Expanding beyond single object texture generation, 
InstanceTex~\cite{Yang:2024:Instancetex} focuses on texture generation for 3D scenes, employing a local synchronized multi-view diffusion strategy to improve local texture consistency across multiple objects. 
3D Paintbrush~\cite{Cecatur:2024:Paintbrush} 
 specializes in localized stylization of single objects, using cascaded score distillation to refine textures within specific object regions. These approaches differ from our 
method in scope: InstanceTex is tailored for stylistic consistency in large environments, while 3D Paintbrush targets localized edits.

%%%% Method

\begin{figure*}[!t]
    \includegraphics[width=\textwidth]{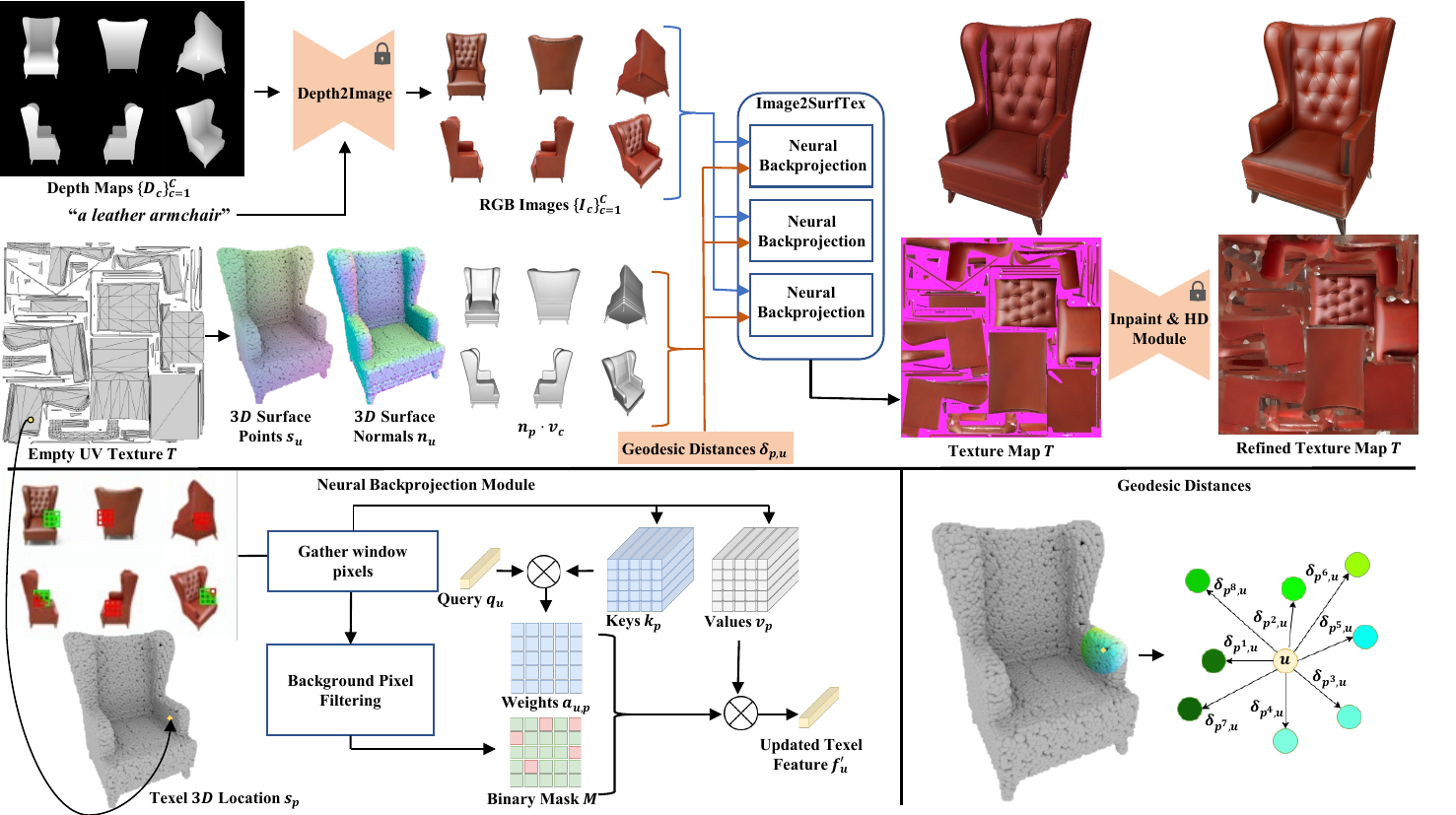}
    \vspace*{-2mm}
    \caption{(Top) The\method, pipeline utilizes depth images and a text prompt to generate a number of candidate views (RGB images) for a given shape. The views are aggregated through a learned backprojection module that incorporates geometric information, such as 3D location, normals, angles between normals, and view vectors, as well as geodesic neighborhood information (bottom right) of shape points corresponding to pixels of the generated RGB images.
 The backprojection module integrates several cross-attention blocks
 (bottom left) used
 to infer texel features and colors from the appearance and geometric information gathered from relevant, non-background pixels across all
 available views.  
 As some texels may remain uncolored, an inpainting and high-definition (HD) module is applied to refine the texture map following Paint3D \cite{Zeng:2024:Paint3D}. }
    \label{fig:architecture}
    \vspace*{-5mm}
\end{figure*}

\section{Method}

Given an untextured 3D shape $\bS$, represented as a polygon mesh, along with its surface parametrization in terms of UV coordinates and a text prompt $\bt$ describing its intended texture,
the goal of our method is to generate an albedo texture map i.e., the base RGB color of the object. The texture map
$\bT$
is 
 stored as a high-res $H \times W \times 3$ atlas in UV space
 ($H=W=1024$ in our implementation). Our overall pipeline is illustrated in Figure \ref{fig:architecture}. Its stages involve: (a) rendering depth maps for the input shape from a set of viewpoints, (b) generating RGB views for these viewpoints through a diffusion model  conditioned on the input depth maps, (c) backprojecting the RGB images to the shape's texture space, (d) inpainting and upsampling the map in UV space. 
  In the following sections, we discuss the steps of our pipeline, and in particular the learned backprojection stage, which is our main contribution. 

\subsection{Depth/edge map rendering \& viewpoint selection} 
\label{subsec:backbones}
As done in several recent texture generation approaches \cite{Zeng:2024:Paint3D, Liu:2024:SyncMVD, Ceylan:2024:Matatlas, Cheng:2024:MVPaint}, the first step in our pipeline is to render the mesh into a set of depth maps
$\{\bD_c\}_{c=1}^C$ from various viewpoints,
where $C$ is the total number of viewpoints. These maps are used as conditioning to guide the diffusion process to generate images consistent with the depth cues.
There have been various strategies for viewpoint selection and diffusion model conditioning -- in our paper, we experimented with two backbones: one based on Paint3D \cite{Zeng:2024:Paint3D}, and another based on MatAtlas \cite{Ceylan:2024:Matatlas}, briefly described below. 

\paragraph{Paint3D backbone}
 Paint3D follows an iterative strategy of viewpoint selection, image generation, and backprojection of the generated images to texture space.  
  First, a couple of $1024 \times 512$ depth images are generated from the frontal and rear views of the shape and are concatenated in a $2\times1$ grid. The grid is passed as input to a diffusion process that generates a corresponding $2\times1$ grid of RGB images. The use of both views as input to the diffusion model
  helps with the view consistency \cite{Zeng:2024:Paint3D}.
  The generated images are backprojected to the shape texture through a simple inverse UV mapping strategy -- we discuss backprojection strategies, including ours, in Section \ref{sec:backprojection}. 
The next iteration proceeds with two side-wise viewpoints, from which both depth images and partially colored RGB images are rendered from the partially textured mesh. These are provided as a grid to another diffusion process, whose generated images are again backprojected to UV space. The process repeats for one more step where two other top- and bottom-wise viewpoints are used. In total, three iterations (total $C=6$ views), 
with two viewpoints processed at a time, yielded the best results in Paint3D. Our experiments with this backbone follow the same iterative procedure and viewpoints -- we only modify the backprojection. 

\paragraph{MatAtlas backbone}
MatAtlas \cite{Ceylan:2024:Matatlas} follows a different viewpoint selection, diffusion conditioning,
 and view generation strategy. Initially, a set  
 $400  \times 400$ depth maps are rendered from viewpoints  uniformly sampled from the viewing sphere, and are arranged in a $4 \times 4$ grid.
  In addition, $16$ edge maps are created using the shape's occluding and suggestive contours and are placed also in a grid. These two grids are used as input to the diffusion process that generates a $4 \times 4$ grid of RGB images. These are backprojected and blended into the shape's texture space (discussed in Section \ref{sec:backprojection}). The resulting partially textured shape is rendered from the same viewpoints, and the rendered RGB images along with added partial noise, are passed to a second diffusion process yielding an updated set of RGB views. These are backprojected to the shape's texture space, yielding a sharper texture \cite{Ceylan:2024:Matatlas}. In a third step, additional viewpoints are selected accessing shape regions not textured yet. The textured shape is rendered from these viewpoints and the rendered images are arranged in a grid processed through another diffusion process, which generates another set of RGB images. These are again backprojected to the final shape's texture. In our implementation, we use $6$ initial viewpoints to render depth maps at resolution $512 \times 512$, arranged in a $3 \times 2$ grid (we do not make use of edge maps). The used viewpoints are the same as the ones used in Paint3D for more fair comparisons across the two backbones, and also because we observed that texture details are better preserved from the higher resolution
  depth maps. We replace the MatAtlas backprojection with ours -- the rest of the pipeline follows
  MatAtlas.
  
\subsection{View generation}
Both backbones use a text-to-image  stable diffusion model \cite{Rombach:2022:High} to generate candidate RGB images based on the input grids.
The stable diffusion model gradually denoises a random normal noise image in latent space
$\bz \in \real^{h\times w \times l}$, where ${h=w=64}$ and $l=4$ are the stable diffusion's latent space dimensions. The outputs of the diffusion model blocks are modulated by a ControlNet network branch \cite{Zhang:2023:ControlNet}, which is conditioned on the encoded text, depth map grid, and, depending on the specific backbone and iteration, on the rendered maps derived from partially textured shapes.
The denoised latent is decoded into a grid of images $\{\bI_c\}_{c=1}^C$ for the selected viewpoints:
\begin{equation}
\{\bI_c\}_{c=1}^C = \mD( \bz, \bt, \{\bD_c\}_{c=1}^C, \{\bG_c\}_{c=1}^C
; \tau_{t}, \tau_{d}, \tau_{g})
\end{equation}
where $\bz$ are noisy latents, $\bt$ is the input text, $\{\bD_c\}_{c=1}^C$ are depth maps, $\{\bG_c\}_{c=1}^C$ are rendered images from the partially textured shape (used in Paint3D and MatAtlas after the first iteration), 
$\tau_{t}, \tau_{d}, \tau_{g}$ are encoder networks that produced text, depth, and image representations used as control guidance for the diffusion process.

\subsection{Backprojection}
\label{sec:backprojection}

The goal of the backprojection is to transfer the generated image colors from all used viewpoints 
back to the shape's texture map. We first describe how backprojection has been implemented in previous methods, then we discuss our neural approach. 

\subsubsection{Traditional backprojection}
\paragraph{Inverse UV mapping}
Previous methods use an inverse UV mapping procedure for backprojection. 
Specifically, given each texel in the texture map $\bu=(u,v) \in \bT$, its corresponding 3D surface point $\bs_{\bu}=(x_{\bu},y_{\bu},z_{\bu}) \in \bS$ is first estimated. Practically, this can be implemented by rendering a flattened version of the input polygon mesh $\bS$ with its vertex coordinates replaced with its texture coordinates. Then for each rendered pixel, its barycentric coordinates are calculated within the flattened triangle it belongs to. These are used to interpolate the 3D vertex positions of this triangle in the original mesh to acquire the corresponding 3D point $\bs_{\bu}$ for that texel. 
The procedure assumes that each texture coordinate maps to a single 3D face -- if a texture coordinate is re-used by multiple faces, the texture can be unwrapped to avoid this \cite{Levy:2002:LSCM}.

\paragraph{Backprojection via most front-facing view}
Most previous methods, such as Paint3D \cite{Zeng:2024:Paint3D}, Text2Tex \cite{Chen:2023:Text2tex}, 
TEXTure~\cite{Richardson:2023:Texture}, 
 find the view where the 3D point appears to be the most front-facing i.e., the dot product between its normal $\bn_{\bu}$ and the view vector $\bv_c$ is maximized,  and simply copy the color from the generated image pixel where the 3D point is projected onto under that view:
\begin{equation}
 \bT[\bu] = \bI_{c'}[ \mR_{c'}( \bs_{\bu} ) ], \;    \text{where}  \;
 c' = argmax_c \big( \bn_{\bu} \cdot \bv_c \big)
\label{eq:backprojection_paint3d}
\end{equation}
where $\mR_{c'}$ returns the 2D pixel coordinates of the point $\bs_{\bu}$ rendered onto the image $\bI_{c'}$ under the most front-facing viewpoint $c'$ for this point. It is also common to employ a hand-tuned threshold 
$\bn_{\bu} \cdot \bv_c > thr$ to avoid copying colors from  obscure views. We note some texels may not acquire any color, if their corresponding points are not accessible by any acceptable views -- texture inpainting is used to fill such texels with color \cite{Zeng:2024:Paint3D}. Unfortunately, this strategy can easily lead to inconsistencies e.g., texels of neighboring 3D points might acquire colors from different views that may not blend well together. 

\paragraph{Backprojection via blending views}
An alternative strategy, followed by MatAtlas \cite{Ceylan:2024:Matatlas} in its first diffusion iteration, 
is to average colors from the pixels of all views accessing the texel's corresponding point to blends any small inconsistencies:
\begin{equation}
 \bT[\bu] = \avg_c \, \bI_{c}[ \mR_{c}( \bs_{\bu} ) ] , 
 \label{eq:backprojection_matatlas}
\end{equation}
Other approaches \cite{zhang2024texpainter, Cheng:2024:MVPaint} implement a weighted averaging scheme, where the weights are the dot product between the 3D point normals $\bn_{\bu}$ and view vectors $\bv_c$.
Unfortunately, averaging schemes can yield blurry texture results, as also noted in \cite{Ceylan:2024:Matatlas}.

\subsubsection{Neural backprojection} 

Instead of relying on ad hoc, hand-tuned schemes for backprojecting and blending colors from
the generated views, we instead propose a learned backprojection scheme. We utilize a neural 
module based on attention \cite{Vaswani:2017:Attenion} to assign appropriate colors to each 
texel by comparing its features with those of pixels gathered from image neighborhoods 
related to this texel across all views. The texel and pixels features are learned based on   positional encodings of the underlying 3D points corresponding to these texels and pixels
 as well as their underlying appearance (color). The positional encodings incorporate information about their 3D position, normals, angles between normals and view vectors, and surface coordinates encoded in geodesic distances -- the reason for using all this information is that the texel color should not be determined by a pixel from a single view, or by merely averaging pixels, but instead by considering
 broader pixel neighborhoods across all views to maximize view consistency,
 and by   
 considering texture correlations in local surface neighborhoods according to the underlying 3D geometry
 to promote texture consistency.

\paragraph{Pixel neighborhoods}
For each texel $\bu$ and each input view, we collect the $K \times K$ pixel neighborhood
centered around the pixel $\mR_{c}( \bs_{\bu} )$, where the texel's corresponding point $\bs_{\bu}$ is projected onto. 
We discard any pixels that lie outside the shape's silhouette, i.e., those in the background. 
The remaining pixels from neighborhoods across all views are then gathered to form a set of pixels
$\mN(\bs_{\bu})$.
The features from these pixels are used as input to our neural module, which learns to determine the texel's color by identifying relevant pixels from this set.
We discuss the choice of $K$ in our experiments.
 While one could theoretically use a very large $K$ (even the entire image), this would be inefficient and degrade performance. 
 Limiting $K$ to $1$, which only includes pixels where the 3D point projects, results in less view-consistent textures in our experiments. 
  We found that smaller neighborhoods ($K=3$) yield the most consistent textures.
 
\paragraph{Positional encodings}
For each pixel $\bp \in \mN(\bs_u)$, we determine the corresponding 3D surface point projected onto this pixel based on the view the pixel originated from. 
We then compute a feature vector that encodes the 3D position $\bs_{\bp}$ and normal $\bn_{\bp}$ of this surface point relative to the texel's corresponding surface point.
Pixels whose 3D locations are closer to the texel's point, or have more similar normals, are expected to have a stronger influence on its color.
Additionally, we encode the geodesic distance $\delta_{\bp, \bu}$ between the pixel's surface point and the texel's 3D point.
Geodesic distances refine pixel contributions by accounting for true surface proximity unlike Euclidean distances, which may misleadingly suggest closeness e.g., in regions with folds and high-curvature regions (Figure \ref{fig:surfconsistency}). Geodesic distances are computed using the method in \cite{Melvaer:2012:GPC}. The encoding is obtained via a trained MLP using the following features:
\begin{equation}
\bh_{\bp} = MLP( \bs_{\bp} - \bs_{\bu}, \bn_{\bp} - \bn_{\bu}, \bn_{\bp} \cdot \bv_{\bc}, \delta_{\bp, \bu} )
\label{eq:positional_encodings}
\end{equation}
The texel's encoding  $\bh_{\bu}$ is also computed using the same MLP. Since we encode relative positions and normals, the texel itself is represented by zero vectors for position and normal differences, and a geodesic distance of zero. We note that absolute 3D positions and normals are not included in our encodings, as they were found to degrade performance.

\paragraph{Appearance encodings}
The texel color should be determined as a function of the pixel color in the extracted neighborhoods, thus we also encode color features used as input to our backprojection module.
For each pixel $\bp \in \mN(\bs_u)$, we use an MLP  to encode its RGB color into a feature vector $\bff_{\bp}$. The same MLP is used to encode the texel's current color into $\bff_{\bu}$, provided it has been initialized from a previous backprojection step. If the texel is empty, we use black color as the input to the MLP.

\begin{figure*}[!t]
    \includegraphics[width=\textwidth]{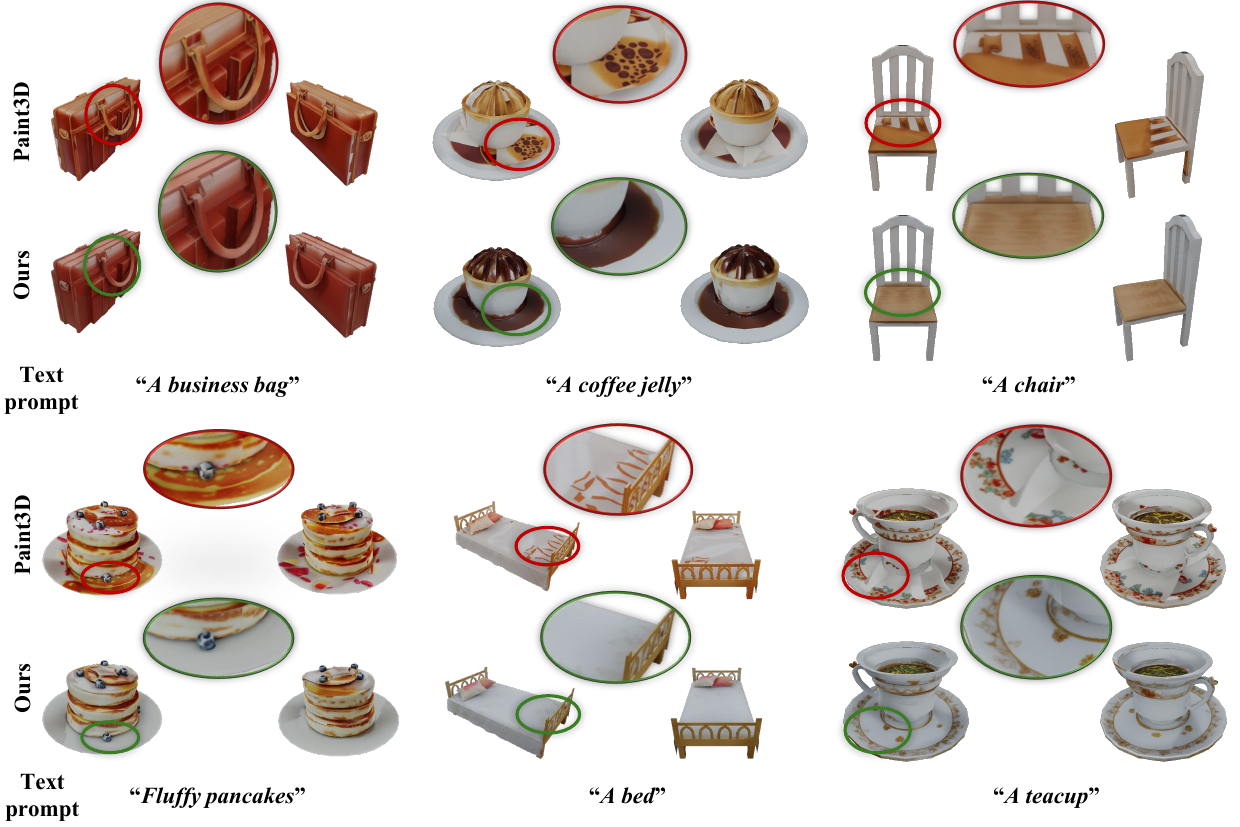}
    \vspace*{-5mm}
    \caption{Comparisons between our method,\method, and Paint3D \cite{Zeng:2024:Paint3D}. Paint3D suffers from view projection artifacts when there are steep depth changes or occluded regions in the input views, as its heuristic best view selection strategy leads to texture discontinuities and inconsistencies. In contrast, our approach generates more seamless and coherent textures.}
    \label{fig:paint3d_comparison}
    \vspace*{-5mm}
\end{figure*}

\paragraph{Cross attention}
To compute the texel color, our module employs a cross-attention mechanism that compares the texel's position and color encoding with those of neighboring pixels to determine their contribution towards the texel color. Specifically, we treat the texel as the query and each pixel as a key, applying the following query-key-value transformations:
\begin{align}
\bq_{\bu} = \bQ \cdot ( \bff_{\bu} + \bh_{\bu} ) \\
\bk_{\bp} = \bK \cdot ( \bff_{\bp} + \bh_{\bp} ) \\
\bv_{\bp} = \bV \cdot  \bff_{\bp} 
\end{align}
where $\bQ, \bK, \bV$ are learned transformations. 
Note that the positional encodings are added to the rest of the features, as also done in 
\cite{Vaswani:2017:Attenion}. Note that in our case, the value transformation involves only the color 
encodings, as our end goal is to transform pixel colors (rather than position) to  texel 
colors. Based on the query and key transformations, we compute the attention weights, which 
represent the importance of each pixel in contributing to the texel's color:
\begin{equation}
a_{\bu, \bp} = softmax( \bq_{\bu} \cdot \bk_{\bp} / \sqrt{D} )
\end{equation}
where $D$ is the dimensionality of the feature vectors ($D=64$ in our implementation). Finally  texel features are updated based on the computed attention weights and a residual block: 
\begin{equation}
\bff_{\bu}' = \sum_{\bp} a_{\bu, \bp} \bv_{\bp} + \bff_{\bu}
\end{equation}

The computed texel features serve as input to a subsequent cross-attention block -- our module applies a total of three attention blocks. The final texel features are then decoded into RGB colors using a trained MLP. Each texel with a non-empty pixel neighborhood is processed through this pipeline. Texels without detected pixel neighborhoods, corresponding to regions inaccessible from any view, remain empty (non-colored); we discuss inpainting for these cases in the next section.

\subsection{Texture inpainting and refinement}
After backprojection and the final iteration of either backbone, some texels may still remain empty. For 
texture inpainting, we follow Paint3D's approach: a trained diffusion model fills any texture holes 
within the UV plane. Additionally, Paint3D's high-definition (HD) diffusion model is subsequently used 
to further enhances the visual quality of the texture map in UV space. We refer readers to Paint3D 
\cite{Zeng:2024:Paint3D} for more details, and the authors' implementation for these trained modules.
We note that we apply the same inpainting and HD processing for both backbone implementations. 
Unfortunately, as shown in our experiments, these post-processing modules often fail to correct the 
artifacts introduced by traditional backprojection.

\subsection{Training}
\label{sec:training}
We train the parameters of our MLPs and cross-attention module based on supervision from 
Objaverse \cite{Deitke:2023:Objaverse}. We use the training split from Paint3D, a subset of 
the Objaverse dataset containing approximately 100K shapes, each paired with a target 
texture image. We preprocess the data by computing geodesic neighborhoods for each object, 
storing the resulting tensors as additional shape-specific information.
The network renders input views, which are then used to reconstruct the target texture 
during training.  Our network is trained with a 
batch size of $4$ for $10$ epochs on four NVIDIA A6000 GPUs, taking approximately five days. 
To make our model more robust to any view inconsistencies,
we employ a mixed batch approach where some renders are re-generated using a pretrained 
Stable Diffusion 1.5 model~\cite{Rombach:2022:High} with partial noise levels ranging from 
$0.2$ to $0.7$. Samples with $0.2$ noise introduce minor variations, while those with $0.7$ 
noise introduce significant deviations from the target texture. For training, we optimize 
the model’s weights using an L1 loss function between the generated and target texture 
images. \vspace*{-4mm}

\subsection{Implementation details} 
During inference, our approach follows either backbone described in Section \ref{subsec:backbones}, yet 
incorporating the learned backprojection module instead of their heuristic backprojection. Since some 
texels remain unfilled after backprojection, they are subsequently inpainted and refined using 
pretrained Stable Diffusion 1.5 and the Paint3D's 3D-aware ControlNet module. The final output textures 
have a resolution of $1024\times1024$. The entire texturing process takes one to two minutes on a single 
NVIDIA A6000 GPU to texture an input shape. Each iteration of view generation and neural backprojection 
takes approximately $35$ seconds, while the inpainting and high-definition (HD) modules require around 
$20$ seconds each. For preprocessing, our approach employs a one-time procedure to compute geodesic 
information metadata, which takes approximately $30$ minutes when processing a new object for the first 
time. This part can be significantly accelerated with more efficient techniques for computation of 
geodesics \cite{Crane:2013:Heat,Zhang:2023:NeuroGF}. We also refer readers to our project page with 
source code for more details. \footnote{\emph{Project page (with code): 
\href{https://ygeorg01.github.io/Im2SurfTex/}{ygeorg01.github.io/Im2SurfTex}}} 

\begin{figure*}[!t]
    \includegraphics[width=\textwidth]{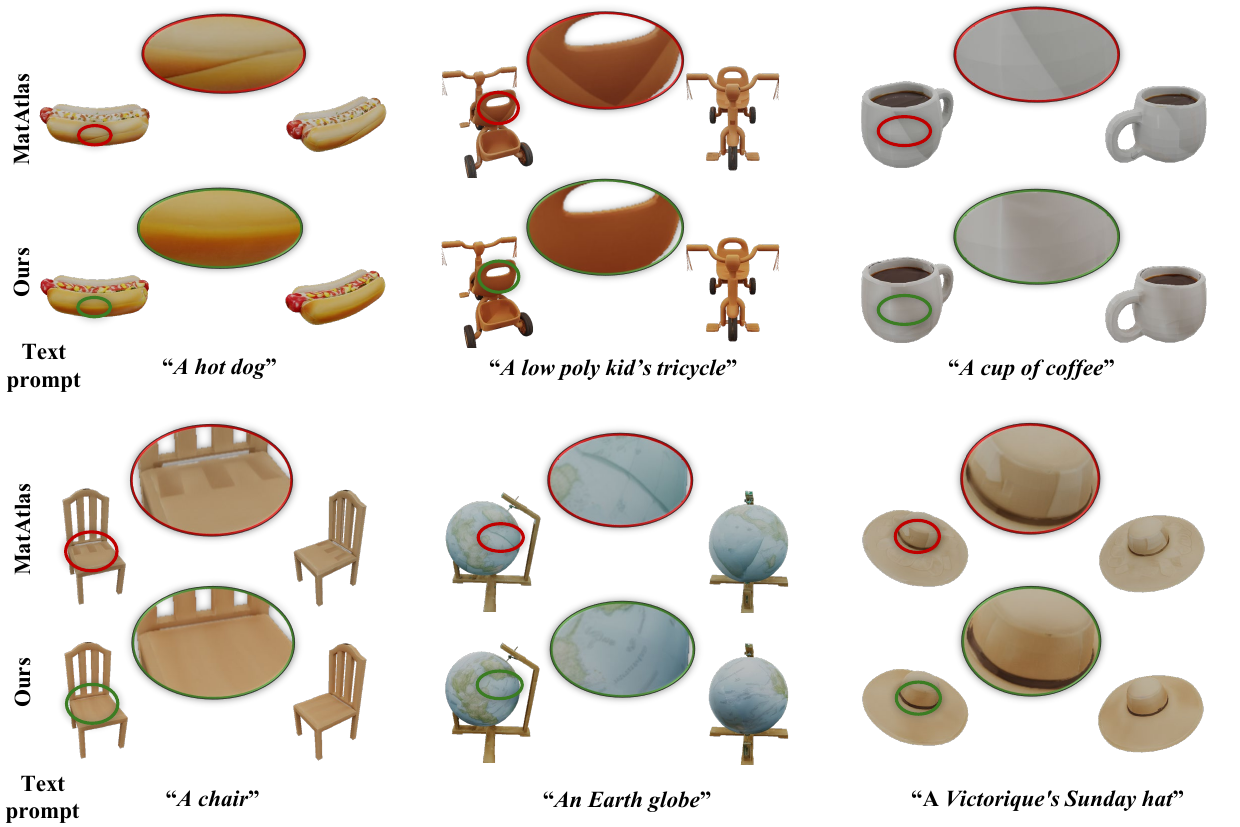}
    \vspace*{-5mm}
    \caption{Comparisons between our method and MatAtlas \cite{Ceylan:2024:Matatlas}. MatAtlas struggles with
    inconsistencies in the output texture, particularly in regions with high curvature, where misalignments 
    become more apparent. In contrast, as shown in the figure, \method\, tends to produce more coherent 
    textures.}
    \label{fig:matatlas_comparison}
    \vspace*{-5mm}
\end{figure*}

%%%% Evaluation
\section{Evaluation}

We evaluate \method \ on text-to-texture generation both quantitatively and qualitatively. In the following sections, we explain the experimental setup for evaluating our approach, including  the evaluation dataset and metrics (Section 
\ref{subsec:experiment_setup}). We then compare \method\,  against competing 
text-to-texture generation methods (Section \ref{subsec:comparisons}). We also analyze the impact 
of pixel neighborhood sizes, geodesic distances and number of input views on performance in an ablation study (Section \ref{subsec:ablation}).

\subsection{Experimental setup}
\label{subsec:experiment_setup}

\paragraph{Test dataset}  
We evaluate \method \ on the test split provided by
Text2Tex~\cite{Chen:2023:Text2tex}. The split includes
$410$ textured meshes from Objaverse~\cite{Deitke:2023:Objaverse} across $225$ categories. All competing methods are trained on the same training split, as discussed in Section \ref{sec:training}, and evaluated on the same above test Objaverse split. We also note that all competing methods use the same UV maps and surface parametrization.

\paragraph{Metrics}
For quantitative evaluation, we use standard image quality metrics for generative image models. Specifically, we report the Fréchet Inception Distance 
(\textbf{FID})~\cite{Heusel:2017:GANs} and Kernel Inception Distance 
(\textbf{KID})~\cite{Bińkowski:2018d:Demystifying}. 
The FID compares the mean and standard deviation of the deepest layer features in the Inception v3 network between the set of real and generated images. The KID calculates the maximum mean discrepancy between the real and generated images. In practice, the MMD is calculated over a number of subsets to obtain a mean and standard deviation measurement.
Additionally, we measure 
alignment, or similarity, of the generated images with the input text prompt using the \textbf{CLIP score}~\cite{Radford:2021:CLIP}. To compute these metrics, following \cite{Zeng:2024:Paint3D}, we 
render each mesh with the generated textures from $20$ fixed viewpoints at a resolution of 
$512\times512$. The reference distribution consists of renders of the same meshes using the textures found 
in the Objaverse dataset, under identical lighting settings.

\subsection{Comparisons}
\label{subsec:comparisons}

Our main finding -- replacing traditional backprojection with our neural module -- is numerically examined in Table~\ref{tab:ablation_backbones}. Our neural module improves both the original Paint3D backbone as well as our implemented MatAtlas backbone. The improvements are consistent across all three evaluation metrics (FID, KID, and CLIP score). Our neural backprojection improves the FID distance by $6.1\%$ for the Paint3D backbone, and $6.9\%$ for the MatAtlas backbone. The improvements are more prominent in terms of the KID score ($19.1\%$ relative reduction for the Paint3D backbone, and $29.2\%$ relative reduction for the MatAtlas backbone). The KID score is more sensitive to fine-grained texture variations due to the use of Maximum Mean Discrepancy (MMD) with a polynomial kernel when comparing distributions. As a result, when a model improvement primarily reduces local inconsistencies -- such as texture artifacts and fine details, KID tends to exhibit a more substantial improvement than FID. In terms of CLIP score, all methods seem to generate images that are similarly aligned with the text prompt, yet our module still maintains a small edge over traditional backprojection.

Figure \ref{fig:paint3d_comparison} and \ref{fig:matatlas_comparison} 
provide comparisons of our module against the Paint3D and MatAtlas respectively. Overall, we observe that our texture results have less artifacts and seams, while preserving a similar level of texture detail. We also refer readers to the supplementary material for more results. 

In Table \ref{tab:comparisons_t2t}, we include quantitative comparisons of the best variant of our method (based on the MatAtlas backbone) with other state-of-the-art models for text-to-texture generation. Here we also include a comparison with the recent method of TEXGen\cite{Yu:2024:TEXGen}.
According to all the evaluation metrics, our method provides the best performance in terms of FID \& KID distances as well as CLIP score.  

\begin{figure}[!t]
    \includegraphics[width=0.48\textwidth]{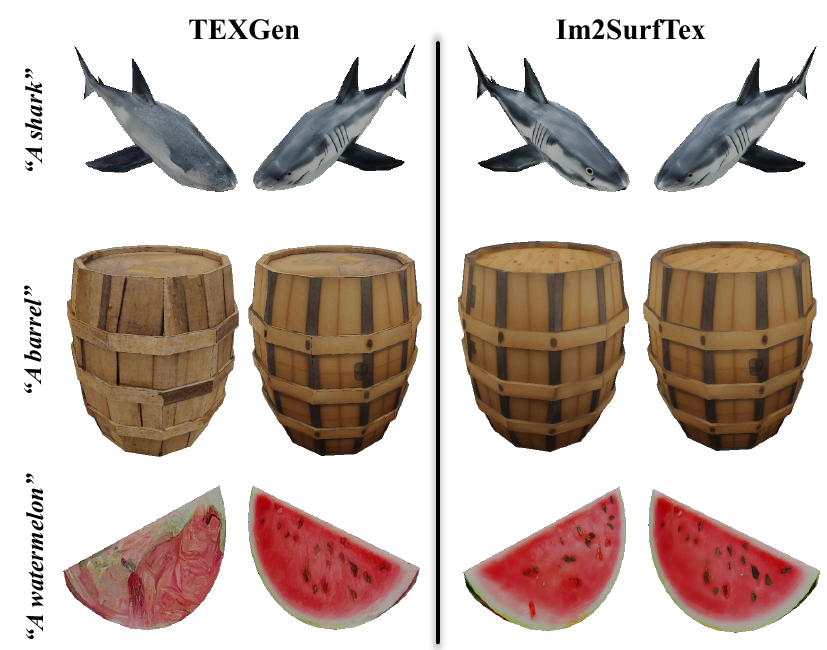}
    \caption{Comparison between \method \ and TEXGen \cite{Yu:2024:TEXGen}. We show  two different views of the textured objects. Our method produces more coherent and view-consistent textures.}
    \label{fig:texgencomparison}
    \vspace*{-4mm}
\end{figure}

In Figure \ref{fig:texgencomparison}, we show qualiitative comparisons with 
TEXGen's released implementation \cite{Yu:2024:TEXGen}. We observe that TEXGen often leads to global texture inconsistencies on the output shapes, while our method is  more view-consistent.

\begin{table}[!t]
    \centering
    \begin{tabular}{c|cccccccc}
        \toprule
         \textbf{Model} & \textbf{FID} $\downarrow$ & \textbf{KID} $\downarrow$ & \textbf{CLIP Score} 
         $\uparrow$  \\ 
         \midrule
        Paint3D & $29.13$ & $2.62$ $\pm$ $0.3$ & $29.45$ \\ 
        Im2SurfTex$_{paint3d}$ & \textbf{$27.34$} & \textbf{$2.12$} $\pm$ \textbf{$0.2$} & \textbf{$29.63$} \\
         \midrule
        MatAtlas & $28.68$ & $2.16$ $\pm$ $0.2$ & $29.65$ \\
        Im2SurfTex$_{matatlas}$ & \textbf{26.68} & \textbf{1.53} $\pm$ \textbf{0.2} & \textbf{29.76} \\
        \bottomrule
    \end{tabular}
    \caption{Evaluation using different backbones for viewpoint selection and image generation. Note that the KID metric includes a mean and standard deviation measurement.}
    \label{tab:ablation_backbones}
    \vspace*{-1mm}
\end{table}

\begin{table}[!t]
    \centering
    \begin{tabular}{c|cccccccc}
        \toprule
         \textbf{Model} & \textbf{FID} $\downarrow$ & \textbf{KID} $\downarrow$ & \textbf{CLIP Score} 
         $\uparrow$ \\ 
         \midrule
        Text2Tex & $34.89$ & $4.82$ $\pm$ $0.3$ & $29.65$  \\ 
        Paint3D & $29.13$ & $2.62$ $\pm$ $0.3$ & $29.45$ &  \\ 
        MatAtlas & $28.68$ & $2.16$ $\pm$ $0.2$ & $29.65$ \\
        TEXGen & $27.41$ & $2.42$ $\pm$ $0.2$ & $29.23$ \\
        Im2SurfTex & \textbf{26.68} & \textbf{1.53} $\pm$ \textbf{0.2} & \textbf{29.76} \\
        \bottomrule
    \end{tabular}
    \caption{Comparisons with other text-to-texture methods.}
    \label{tab:comparisons_t2t}
    \vspace*{-1mm}
\end{table}

\subsection{Ablation}
\label{subsec:ablation}

We provide an ablation study where we vary the input pixel neighborhood size extracted from the generated images
for each texel. Results are shown in  Table~\ref{tab:ablation-neighborhood} for neighborhoods $1 \times 1$ , $3 \times 3$ , $5 \times 5$, and $7 \times 7$. Best performance is achieved under the $3 \times 3$ neighborhood setting. 

\begin{figure}[!t]
    \includegraphics[width=0.45\textwidth]{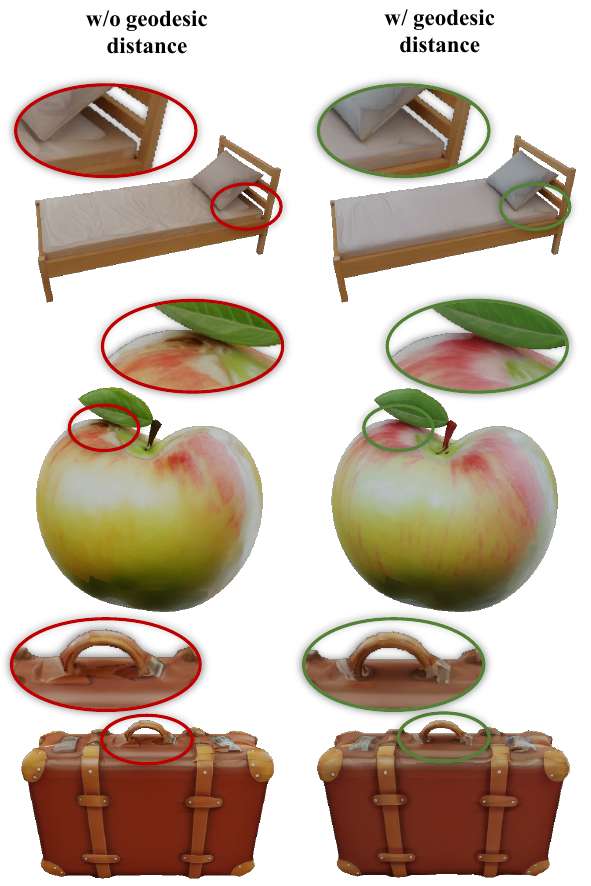}
    \caption{Using geodesic distances in the positional encodings of texels promote texture consistency. On the left, Im2SurfTex operates without geodesic information, resulting in less coherent textures in areas with rapidly changing local geometry
    (e.g., surface regions with folds, handles, or high curvature). On the right, incorporating geodesic information improves texture quality in these regions.}
    \label{fig:surfconsistency}
    \vspace*{-4mm}
\end{figure}

Table \ref{tab:ablation-geodesics} provides another ablation where we compare using absolute versus relative coordinates in the positional encodings of the Eq. \ref{eq:positional_encodings}, and also examine whether using geodesic distance as additional feature in the positional encodings helps. Relative coordinates enhance performance compared to absolute coordinates, as they provide a more effective encoding for processing the local interactions between neighboring points, regardless of their actual 3D locations. With repect to the use of geodesic distances, we observe rather minor improvements in terms of the numerical scores. We suspect that the small differences are due to the fact that the improvements happen only in small image regions for the shapes of our dataset,
where the surface changes rapidly (e.g., folds, handles, high curvature regions), as shown in Figure \ref{fig:surfconsistency}. These small regions seem to have a relatively small effect on the established image quality metrics. 
Figure \ref{fig:surfconsistency} demonstrates  that adding geodesic distances as features in our module leads to fewer texture artifacts and diminished color bleeding in these regions e.g., see the color bleeding between the bed mattress and wooden frame, or the green leaf and the apple when geodedic distances are not used. 

\begin{figure}[!t]
    \includegraphics[width=0.49\textwidth]{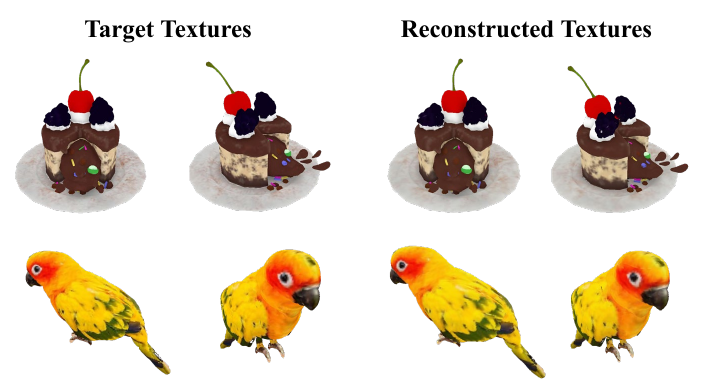}
     \vspace*{-5mm}
    \caption{Our neural backprojection can closely reconstruct challenging textures of target objects in our dataset without  causing noticeable color shifting or discrepancies between target and decoded textures.}
    \label{fig:fig_reconstruction}
    \vspace*{-5mm}
\end{figure}

Figure \ref{fig:fig_reconstruction} demonstrates a visual comparison between reference textured meshes from our dataset, and reconstructed textures by our method, when we pass as input the rendered images from the original textures. This comparison aims to show whether our method causes any significant color shifting or bleeding while aggregating information from different views. We see that demonstrating our method does not introduce any such discrepancies during neural backprojection.

\begin{table}[!t]
\centering
\begin{tabular}{c|cccc}
\toprule
Window size  & FID $\downarrow$ & KID $\downarrow$ & CLIPscore $\uparrow$ \\
\midrule
1 $\times$ 1 & $27.65$    & $2.31$ $\pm$ $0.2$ & $29.59$ \\
3 $\times$ 3 & \textbf{27.35}  & \textbf{2.15} $\pm$ \textbf{0.2}  & \textbf{29.61}\\
5 $\times$ 5 & $27.43$ & $2.26$ $\pm$ $0.2$ & $29.60$  \\
7 $\times$ 7 & $28.12$ & $2.36$ $\pm$ $0.3$ & $29.54$  \\
\bottomrule
\end{tabular}
\caption{Ablation study results wrt texel neighborhood size (no geodesic distances are used in this experiment)}
\label{tab:ablation-neighborhood}
\vspace*{-1mm}
\end{table}

\begin{table}[!t]
\centering
\begin{adjustbox}{width=0.45\textwidth}
\begin{tabular}{cc|ccc}
\toprule
Rel Coords. & Geod. Distances & FID $\downarrow$ & KID $\downarrow$ & CLIPscore $\uparrow$ \\
\midrule
- & -  & 27.86  & 2.32 $\pm$ 0.2  & 29.62\\
\checkmark & -  &  27.35  & 2.15 $\pm$ 0.2  & 29.61\\
\checkmark & \checkmark & \textbf{27.34} & \textbf{2.12} $\pm$ \textbf{0.3} & \textbf{29.63} \\
\bottomrule
\end{tabular}
\end{adjustbox}
\caption{Ablation study results wrt using geodesic distances or not in the cross-attention operation of our backprojection module. Note that this experiment uses $3 \times 3$ pixel neighborhoods.}
\label{tab:ablation-geodesics}
\vspace*{-1mm}
\end{table}

\begin{table}[!t]
\centering
\begin{adjustbox}{width=0.45\textwidth}
\begin{tabular}{cc|ccc}
\toprule
 \# of Views &  Method &  FID $\downarrow$ &  KID $\downarrow$ &  CLIPscore $\uparrow$ \\
\midrule
\multirow{2}{*}{ 4} &  Paint3D &  29.41  &  2.71 $\pm$ 0.3  &  29.40\\
 &  Im2SurfTex &  \textbf{28.51}  &  \textbf{2.42 $\pm$ 0.3}  &  \textbf{29.62}\\
\hline
\multirow{2}{*}{ 6} &  Paint3D &  29.13 &  2.62 $\pm$ 0.2  &  29.45\\
&  Im2SurfTex &  \textbf{27.34}  &  \textbf{2.12 $\pm$ 0.2} &  \textbf{29.63}\\
\hline
\multirow{2}{*}{ 8} &  Paint3D &  29.20  &  2.75 $\pm$ 0.3  &  29.48\\
 &  Im2SurfTex &  \textbf{27.86} &  \textbf{2.32 $\pm$ 0.2}  &  \textbf{29.62}\\

\bottomrule
\end{tabular}
\end{adjustbox}
\caption{ Ablation study results wrt using different number of views.}
\label{tab:ablation-number-views}
\vspace*{-4mm}
\end{table}

\begin{figure}[!t]
    \includegraphics[width=0.45\textwidth]{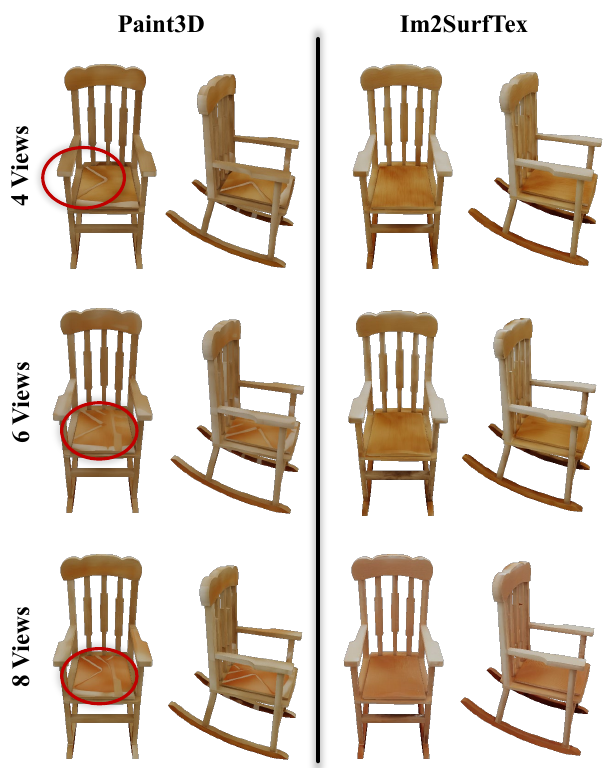}
    \caption{Results for Paint3D and our method for $4$, $6$, and $8$ input views. \method\ generates smoother surfaces.}
    \label{fig:numverofviewsnumber}
    \vspace*{-1mm}
\end{figure}

Table \ref{tab:ablation-number-views} presents the impact of using different number of views on our evaluation metrics for both Paint3D and our method (using the Paint3D backbone). Increasing the number of views from $4$ to $6$ views results in improvements for the FID, KID, and CLIP cores. Yet, for the maximum number of views ($8$) in this experiments, we see that the FID and KID scores do not further improve. As shown in Figure \ref{fig:numverofviewsnumber}, we observe more artifacts appearing in Paint3D, probably due to its use of mere backprojection which often leads to more seams when more views are backprojected. Our method scores with $8$ views are affected less; our neural backprojection produces smoother results, yet we do notice a bit more oversmoothing in our case, which is a limitation our our method.

%%%% Evaluation
\section{Conclusion \& Future Work}

In conclusion, Im2SurfTex presents a novel backprojection approach to generating high-quality, coherent textures for 3D shapes from multiview image outputs from pretrained 2D diffusion models. Unlike conventional methods that rely on heuristic and averaging backprojection strategies that introduce texture artifacts and seams, our approach enhances texture continuity and coherence. Experimental results validate the effectiveness of our method. 

\paragraph{Limitations and future work}  In our current implementation, texture generation is limited by predefined viewpoints that may be suboptimal. Instead, a better approach would be to dynamically adapt to the shape’s intrinsic structure.
Future work can focus on integrating richer geometric information and utilizing specialized 3D networks to encode complex features such as curvature and occluded regions,
which remain challenging for current approaches. By enabling geometry-aware processing in the diffusion process, future methods may further mitigate view-dependent biases.

\paragraph{Acknowledgements} This project has received funding from the European Research Council (ERC) under the Horizon Research and Innovation Programme (Grant agreement No. 101124742). Additionally, it has been supported from the EU H2020 Research and Innovation Programme and the Republic of Cyprus through the Deputy Ministry of Research, Innovation and Digital Policy (Grant agreement No. 739578).

%-------------------------------------------------------------------------
% bibtex
\bibliographystyle{eg-alpha-doi} 
\bibliography{bibliography}          

\end{document}